\documentclass[final,5p,times,twocolumn]{elsarticle}

\usepackage{amsmath}
\usepackage{amsfonts} 
\usepackage{mathtools} 
\usepackage{amssymb} 
\usepackage{subfig} 
\usepackage{xcolor}
\usepackage{mathrsfs}
\usepackage[breaklinks,colorlinks,urlcolor=blue,citecolor=blue,linkcolor=magenta]{hyperref}
\usepackage{cleveref}
\usepackage{multirow}
\usepackage{orcidlink}
\usepackage{comment}
\usepackage{etoolbox}
\usepackage{stfloats} 
\usepackage{cuted}    

\biboptions{sort&compress}

\newcommand{\abs}[1]{|#1|} 
\newcommand{\re}[1]{\text{Re}\left(#1\right)}
\newcommand{\im}[1]{\text{Im}\left(#1\right)}

\newcommand{\SD}[1]{\Phi_{#1}^{\phantom{\dagger}}}
\newcommand{\SDd}[1]{\Phi_{#1}^\dagger}

\newcommand{\SDp}[1]{\Phi_{#1}^{\prime\phantom{\dagger}}}
\newcommand{\SDpd}[1]{\Phi_{#1}^{\prime\dagger}}

\newcommand{\qq}[1]{\mu^2_{#1}}
\newcommand{\qqc}[1]{\mu^{2\,\ast}_{#1}}
\newcommand{\QQ}[1]{\lambda_{#1}}
\newcommand{\QQc}[1]{\lambda_{#1}^\ast}
\newcommand{\nfR}[1]{\mathrm{R}_{#1}}
\newcommand{\nfI}[1]{\mathrm{I}_{#1}}

\newcommand{\chfP}[1]{\mathrm{C}^+_{#1}}
\newcommand{\chfM}[1]{\mathrm{C}^-_{#1}}
\newcommand{\chfvec}{{\vec{\mathbf{C}}}}
\newcommand{\nfvec}{{\vec{\mathbf{N}}}}

\newcommand{\FchfPM}[1]{\mathrm{H}^\pm_{#1}}

\newcommand{\FchfM}[1]{\mathrm{H}^-_{#1}}
\newcommand{\Fnf}[1]{\mathrm{H}_{#1}^0}
\newcommand{\Omat}{O}\newcommand{\OmatT}{\Omat^T}

\newcommand{\Umat}{U}\newcommand{\UmatDag}{\Umat^\dagger}

\newcommand{\VEV}[1]{\langle #1 \rangle}
\newcommand{\vev}[1]{v_{#1}}
\newcommand{\vevPh}[1]{\theta_{#1}}
\newcommand{\nMM}{{\rm M_0^2}}
\newcommand{\chMM}{{\rm M_{\pm}^2}}
\newcommand{\potV}{\mathcal{V}}
\newcommand{\vevV}{\mathrm{V}}
\newcommand{\TR}[1]{\text{Tr}\left(#1\right)}

\def\eg{\textit{e.g.}}

\def\vs{\textit{vs.}}

\graphicspath{{NewFigs/}{Figs/}}

\journal{Physics Letters B}

\begin{document}

\begin{frontmatter}

\title{On vacua and bounded masses in the general 2HDM}

\author[ific]{Jos\'e M. Camacho\,\orcidlink{0009-0001-8856-258X}}
\ead{jose.m.camacho@uv.es}

\author[infn]{Carlos Mir\'o\,\orcidlink{0000-0003-0336-9025}}
\ead{carlos.miroarenas@to.infn.it}

\author[ific]{Miguel Nebot\,\orcidlink{0000-0001-9292-7855}}
\ead{Miguel.Nebot@uv.es}

\author[ific]{Tom\'as Tobarra\,\orcidlink{0009-0000-6773-4023}}
\ead{Tomas.Tobarra@ific.uv.es}

\affiliation[ific]{organization={Instituto de F\'isica Corpuscular (IFIC), CSIC-UV \& Departament de F\'isica Te\`orica, Universitat de Val\`encia},
            city={Valencia},
            postcode={E-46100},
            country={Spain}}

\affiliation[infn]{organization={INFN, Sezione di Torino},
            addressline={Via Pietro Giuria 1},
            city={Turin},
            postcode={I-10125},
            country={Italy}}

\begin{abstract}
Two Higgs doublets models with a scalar potential that breaks electroweak symmetry spontaneously can have either one or two local minima. While potentials with one minimum can have a decoupling regime where all the new scalars are heavy, we show that, for potentials with \emph{two} local minima, the masses of all the scalars are bounded if the dimensionless quartic couplings obey perturbativity constraints.
\end{abstract}


\end{frontmatter}

\section{Introduction\label{SEC:Intro}}
Knowing the masses of new particles arising in scenarios beyond the Standard Model (SM) is of great importance. Apart from direct or indirect experimental information, theoretical arguments have been put forward to constrain possible mass spectra \cite{Weinberg:1976pe,Politzer:1978ic,Cabibbo:1979ay,Dashen:1983ts,Callaway:1983zd}, including in particular perturbativity requirements as proposed by Lee, Quigg and Thacker \cite{Lee:1977eg,Lee:1977yc,Dicus:1973gbw} to set bounds on the Higgs mass. In extended scalar sectors inducing spontaneous symmetry breaking, these perturbativity requirements imply the presence of a Higgs-like scalar with electroweak-scale mass, set by the spontaneous symmetry breaking vacuum expectation value (vev) $\vev{\rm EW}\simeq 246$ GeV \cite{Weldon:1984th,Langacker:1984dma,Comelli:1996xg,Espinosa:1996vt}. In multi-Higgs models \cite{Ivanov:2017dad}, starting with the minimal scenario with two doublets (2HDM) \cite{Lee:1973iz,Lee:1974jb}, these constraints have been explored, for example, in \cite{Huffel:1980sk,Casalbuoni:1987cz,Maalampi:1991fb,Kanemura:1993hm,Ginzburg:2005dt,Horejsi:2005da,Kanemura:2015ska}. In some cases, not necessarily multi-Higgs scenarios, scalar sectors where an imposed symmetry yields a number of quadratic parameters in the potential not larger than the number of stationarity conditions, the quadratic parameters can be expressed in terms of bounded quartic parameters and vacuum expectation values, giving as a result that all the new scalars have bounded masses. One illustrative case is the 2HDM where the potential is CP invariant and the vacuum breaks CP spontaneously: the 3 real quadratic parameters in the potential are reexpressed in terms of dimensionless quartic parameters and vevs, and then perturbativity constraints on the quartic parameters force $\mathcal O(\vev{\rm EW})$ bounds on all scalar masses \cite{Nebot:2018nqn,Nebot:2019qvr,Nierste:2019fbx}. With more than two doublets, a CP invariant potential and spontaneous CP violation (SCPV), there are free, unconstrained quadratic parameters. Unexpectedly, this freedom cannot produce masses much larger than the electroweak scale for all the new scalars, as discussed in \cite{Miro:2024zka,Camacho:2026hrc}. One critical ingredient appears to be the following: in that context, if there is a CP violating candidate vacuum (a minimum of the potential), its CP-conjugate is also a candidate vacuum (with, in fact, the same depth).

This motivates the present work: since the general 2HDM can have either one or two possible vacua \cite{Ivanov:2007de}, is something similar at work? 2HDMs, or at least some of them, can have a decoupling regime for the new scalars \cite{Haber:1989xc,Gunion:2002zf} (see \cite{Das:2026ldf} for recent work); the goal here is to provide a different perspective: what can be said, concerning the mass spectrum, depending on the fact that the scalar potential has either one or two local minima? The surprising answer is that in the case of a potential with two minima, the masses of all the scalars are bounded, they cannot be much larger than the electroweak scale, if perturbativity constraints are imposed on the quartic couplings.

The manuscript is organized as follows. In \cref{SEC:2HDM} we set notation and address some basic aspects. A detailed numerical analysis is presented in \cref{SEC:NumPheno}. In \cref{SEC:Discussion} we discuss the central results, and then present our conclusions.

\section{The general 2HDM\label{SEC:2HDM}}
The most general scalar potential for two $SU(2)_L$ doublets with hypercharge $1/2$, $\SD{1}$  and $\SD{2}$, is
\begin{equation}\label{eq:2HDM:V:01}
\begin{aligned}
\potV(\SD{1},\SD{2}) &= \potV_2(\SD{1},\SD{2})+\potV_4(\SD{1},\SD{2})\,,\\[5pt]
\potV_2(\SD{1},\SD{2}) &= \qq{1}\,\SDd{1}\SD{1}+\qq{2}\,\SDd{2}\SD{2}+\qq{12}\,\SDd{1}\SD{2}+\qqc{12}\,\SDd{2}\SD{1}\,,\\[5pt]
\potV_4(\SD{1},\SD{2}) &=
\QQ{1}(\SDd{1}\SD{1})^2+\QQ{2}(\SDd{2}\SD{2})^2+\QQ{3}(\SDd{1}\SD{1})(\SDd{2}\SD{2})\\
&\ +\QQ{4}(\SDd{1}\SD{2})(\SDd{2}\SD{1})
+\QQ{5}(\SDd{1}\SD{2})^2+\QQc{5}(\SDd{2}\SD{1})^2\\
&\ +(\QQ{6}\,\SDd{1}\SD{2}+\QQc{6}\,\SDd{2}\SD{1})(\SDd{1}\SD{1})\\
&\ +(\QQ{7}\,\SDd{1}\SD{2}+\QQc{7}\,\SDd{2}\SD{1})(\SDd{2}\SD{2})\,,
\end{aligned}
\end{equation}
where $\qq{1}$, $\qq{2}$, $\QQ{1}$, $\QQ{2}$, $\QQ{3}$, $\QQ{4}$ are necessarily real, while $\qq{12}$, $\QQ{5}$, $\QQ{6}$, $\QQ{7}$ are, in general, complex. Spontaneous symmetry breaking occurs for a minimum of the function
\begin{equation}
 \vevV(\VEV{\SD{1}},\VEV{\SD{2}})\equiv\potV(\VEV{\SD{1}},\VEV{\SD{2}})\,,
\end{equation}
with $\VEV{\SD{a}}\neq\left(\begin{smallmatrix}0\\ 0\end{smallmatrix}\right)$. Electroweak spontaneous symmetry breaking is assumed for vevs
\begin{equation}\label{eq:vevs:01}
\langle\SD{a}\rangle=\frac{\vev{a}e^{i\vevPh{a}}}{\sqrt 2}\begin{pmatrix}0\\ 1\end{pmatrix},\quad \vev{a},\vevPh{a}\in\mathbb{R}\,,\quad \vev{a}\geq 0\,,
\end{equation}
and the fields are expanded as
\begin{equation}\label{eq:FieldExp:01}
 \SD{a}=\frac{e^{i\vevPh{a}}}{\sqrt 2}\begin{pmatrix}\sqrt{2}\chfP{a}\\ \vev{a}+\nfR{a}+i\,\nfI{a}\end{pmatrix}.
\end{equation}
Only the phase difference $\vevPh{}\equiv \vevPh{2}-\vevPh{1}$ (mod $2\pi$) is physical and can produce spontaneous CP violation. We use $\vev{}\equiv\sqrt{\vev{1}^2+\vev{2}^2}$; for adequate electroweak symmetry breaking, one sets $\vev{}=\vev{\rm EW}$. The 3 stationarity conditions, necessarily fulfilled at a minimum of the potential, are
\begin{align}
\label{eq:statcond:v1}
\partial_{\vev{1}}\vevV =& \qq{1}\vev{1}+\re{\qq{12}e^{i\vevPh{}}}\vev{2}+\QQ{1}\vev{1}^3+\frac{1}{2}(\QQ{3}+\QQ{4})\vev{1}\vev{2}^2\nonumber\\
&+\re{\QQ{5}e^{i2\vevPh{}}}\vev{1}\vev{2}^2 +\frac{3}{2}\re{\QQ{6}e^{i\vevPh{}}}\vev{1}^2\vev{2}+\frac{1}{2}\re{\QQ{7}e^{i\vevPh{}}}\vev{2}^3\nonumber\\
&= 0 \,,\\[5pt]
\label{eq:statcond:v2}
\partial_{\vev{2}}\vevV =& \qq{2}\vev{2}+\re{\qq{12}e^{i\vevPh{}}}\vev{1}+\QQ{2}\vev{2}^3+\frac{1}{2}(\QQ{3}+\QQ{4})\vev{1}^2\vev{2}\nonumber\\
&+\re{\QQ{5}e^{i2\vevPh{}}}\vev{1}^2\vev{2}+\frac{1}{2}\re{\QQ{6}e^{i\vevPh{}}}\vev{1}^3+\frac{3}{2}\re{\QQ{7}e^{i\vevPh{}}}\vev{1}\vev{2}^2\nonumber\\
&=0 \,,\\[5pt]
\label{eq:statcond:th}
\partial_{\vevPh{}}\vevV =& -\im{\qq{12}e^{i\vevPh{}}}\vev{1}\vev{2}-\im{\QQ{5}e^{i2\vevPh{}}}\vev{1}^2\vev{2}^2\nonumber\\
&\ -\im{(\QQ{6}\vev{1}^2+\QQ{7}\vev{2}^2)e^{i\vevPh{}}}\frac{\vev{1}\vev{2}}{2}=0
\end{align}

Notice that, with 4 quadratic parameters and 3 stationarity conditions, one cannot \emph{engineer} a potential with the chosen vevs and all quadratic parameters fixed in terms of the vevs and other parameters: one typically chooses the vacuum configuration $\{\vev{1},\vev{2}e^{i\vevPh{}}\}$, and fixes $\qq{1}$, $\qq{2}$ and $\re{\qq{12}}$ through \cref{eq:statcond:v1,eq:statcond:v2,eq:statcond:th}, while $\im{\qq{12}}$ is free. It is worth stressing, however, that this usual counting argument is questionable since, with a change of $\{\SD{1},\SD{2}\}$ basis, one can remove one or two independent quadratic terms from the scalar potential: a related discussion is postponed to section \ref{SEC:Discussion}.

The mass terms for the charged $\chfvec^\dagger=(\chfP{1},\chfP{2})$ and neutral fields $\nfvec^T=(\nfR{1},\nfR{2},\nfI{1},\nfI{2})$ are
\begin{equation}\label{eq:pot:massterms:01}
 -\mathscr L_{\mathrm{Mass}}=\chfvec^\dagger\,\chMM\,\chfvec+\frac{1}{2}\nfvec^T\,\nMM\,\nfvec\,,
\end{equation}
where $\chMM$ is the $2\times 2$ hermitian charged mass matrix and $\nMM$ is the $4\times 4$ real symmetric neutral mass matrix. $\chMM$ and $\nMM$ have null eigenvalues corresponding to the would-be Goldstone bosons, with eigenvectors
\begin{equation}
\begin{aligned}
&\vec c_G=(\vev{1},\vev{2})^T\,,\\
&\quad\chMM\,\vec c_G=\left(\partial_{\vev{1}}\vevV+\frac{i}{\vev{1}}\partial_{\vevPh{}}\vevV,\ \partial_{\vev{2}}\vevV-\frac{i}{\vev{2}}\partial_{\vevPh{}}\vevV\right)^T=\vec{0}^T\,,\\[5pt]
&\vec n_G=(0,0,\vev{1},\vev{2})^T\,,\\
&\quad\nMM\,\vec n_G=\left(\frac{1}{\vev{1}}\partial_{\vevPh{}}\vevV,\ -\frac{1}{\vev{2}}\partial_{\vevPh{}}\vevV,\ \partial_{\vev{1}}\vevV,\ \partial_{\vev{2}}\vevV\right)^T=\vec{0}^T\,.
\end{aligned}
\end{equation}
$\chMM$ is diagonalized with a $2\times 2$ unitary matrix $\Umat$, which gives the would-be Goldstone bosons $G^\pm$ and the physical charged scalar $\FchfPM{}$:
\begin{equation}
(\chfM{1},\chfM{2})^T=\Umat\,(G^-,\FchfM{})^T\,,\quad \UmatDag\,\chMM\,\Umat=\text{Diag}(0,M_{\FchfPM{}}^2)\,.
\end{equation}
$\nMM$ is diagonalized with a $4\times 4$ real orthogonal matrix $\Omat$, which gives the would-be Goldstone boson $G^0$ and the physical neutral scalars. We assume that the lightest neutral scalar is ``the Higgs'' $h$ (this is just a convenient choice). We have
\begin{equation}
\begin{aligned}
(\nfR{1},\nfR{2},\nfI{1},\nfI{2})^T &=\Omat\,(G^0,h,\Fnf{1},\Fnf{2})^T\,,\\[5pt]
\OmatT\,\nMM\,\Omat &=\text{Diag}(0,m_h^2,M_{\Fnf{1}}^2,M_{\Fnf{2}}^2)\,,
\end{aligned}
\end{equation}
with $m_h\leq M_{\Fnf{1}}\leq M_{\Fnf{2}}$.

The diagonalization of $\chMM$ is straightforwardly achieved by rotating to ``the Higgs basis'' \cite{Botella:1994cs} with
\begin{equation}\label{eq:CHMass:HB}
 \Umat=\frac{1}{\vev{}}\begin{pmatrix}\vev{1} & -\vev{2}\\ \vev{2} & \vev{1}\end{pmatrix}\,.
\end{equation}
From the diagonalization, or computing directly $\TR{\chMM}$, one obtains the mass of the charged scalar
\begin{equation}\label{eq:massCH:01}
\begin{aligned}
M_{\FchfPM{}}^2 =&\qq{1}+\qq{2}+\QQ{1}\vev{1}^2+\QQ{2}\vev{2}^2\\
&+\frac{1}{2}\QQ{3}\vev{}^2+\re{(\QQ{6}+\QQ{7})e^{i\vevPh{}}}\vev{1}\vev{2}\,,
\end{aligned}
\end{equation}
expressed for simplicity in terms of $\qq{1}+\qq{2}$ as the free quadratic parameter left; one can equivalently express $M_{\FchfPM{}}^2$, for example, in terms of $\im{\qq{12}}$:
\begin{equation}
\begin{aligned}
M_{\FchfPM{}}^2 &=\frac{\vev{}^2}{2\vev{1}\vev{2}\sin\vevPh{}}\\[5pt]
&\times \im{2\qq{12}+(2\QQ{5}-\QQ{4})e^{i\vevPh{}}\vev{1}\vev{2}+\QQ{6}\vev{1}^2+\QQ{7}\vev{2}^2}\,.
\end{aligned}
\end{equation}
For $\nMM$, the calculation of the eigenvalues is not straightforward, but it is sufficient for now to notice that, from $\TR{\nMM}$,
\begin{equation}\label{eq:massN:01}
\begin{aligned}
m_h^2+M_{\Fnf{1}}^2+M_{\Fnf{2}}^2 &=
2(\qq{1}+\qq{2})+4(\QQ{1}\vev{1}^2+\QQ{2}\vev{2}^2)\\
&+(\QQ{3}+\QQ{4})\vev{}^2+4\re{(\QQ{6}+\QQ{7})e^{i\vevPh{}}}\vev{1}\vev{2}\,.
\end{aligned}
\end{equation}
From \cref{eq:massCH:01,eq:massN:01}, it is clear that giving masses to the new scalars much larger than the electroweak scale $\vev{\rm EW}$, requires $\qq{1}+\qq{2}\gg \vev{\rm EW}^2$. The sole use of the stationarity conditions in \cref{eq:statcond:v1,eq:statcond:v2,eq:statcond:th} does not appear to bar that possibility, and thus a decoupling regime for (all) the new scalars is possible.
A crucial comment is in order. We have mentioned \emph{the} vacuum $\{\vev{1},\vev{2}e^{i\vevPh{}}\}$ and have discussed accordingly the stationarity conditions, but it might happen that the scalar potential has not one but \emph{two} local symmetry breaking minima, as analyzed in depth in \cite{Ivanov:2007de} (see also \cite{Barroso:2006pa,Barroso:2007rr,Ivanov:2006yq,Ginzburg:2007jn}). If that is the case, then a second solution $\{\vev{1}^\prime,\vev{2}^\prime e^{i\vevPh{}^\prime}\}$ of \cref{eq:statcond:v1,eq:statcond:v2,eq:statcond:th} exists. Then, the possibility of having a decoupling regime might be jeopardized. With three stationarity conditions from each of the two vacua, we have a number of conditions to be fulfilled in excess of the number of quadratic parameters, the latter can be expressed as products of quartic couplings and vevs, and thus bounded by perturbativity requirements.

\section{Numerical exploration\label{SEC:NumPheno}}
\begin{figure*}[!ht]
\centering
\subfloat[$M_{\Fnf{2}}$ \vs\ $M_{\Fnf{1}}$.\label{sfig:vac1:Masses:M02vsM01}]{\includegraphics[width=0.44\textwidth]{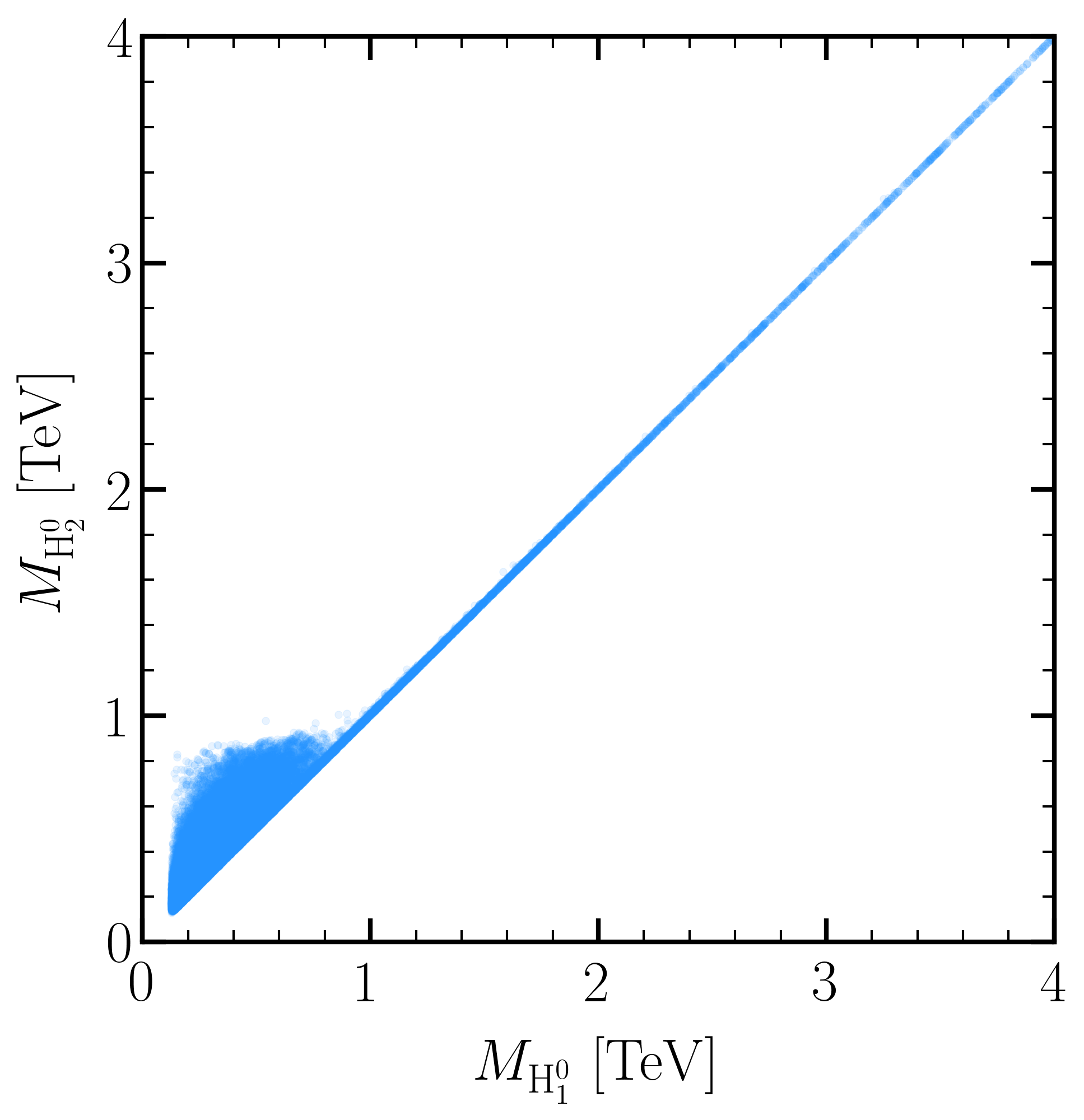}}\qquad
\subfloat[$M_{\FchfPM{1}}$ \vs\ $M_{\Fnf{1}}$.\label{sfig:vac1:Masses:MC1vsM01}]{\includegraphics[width=0.44\textwidth]{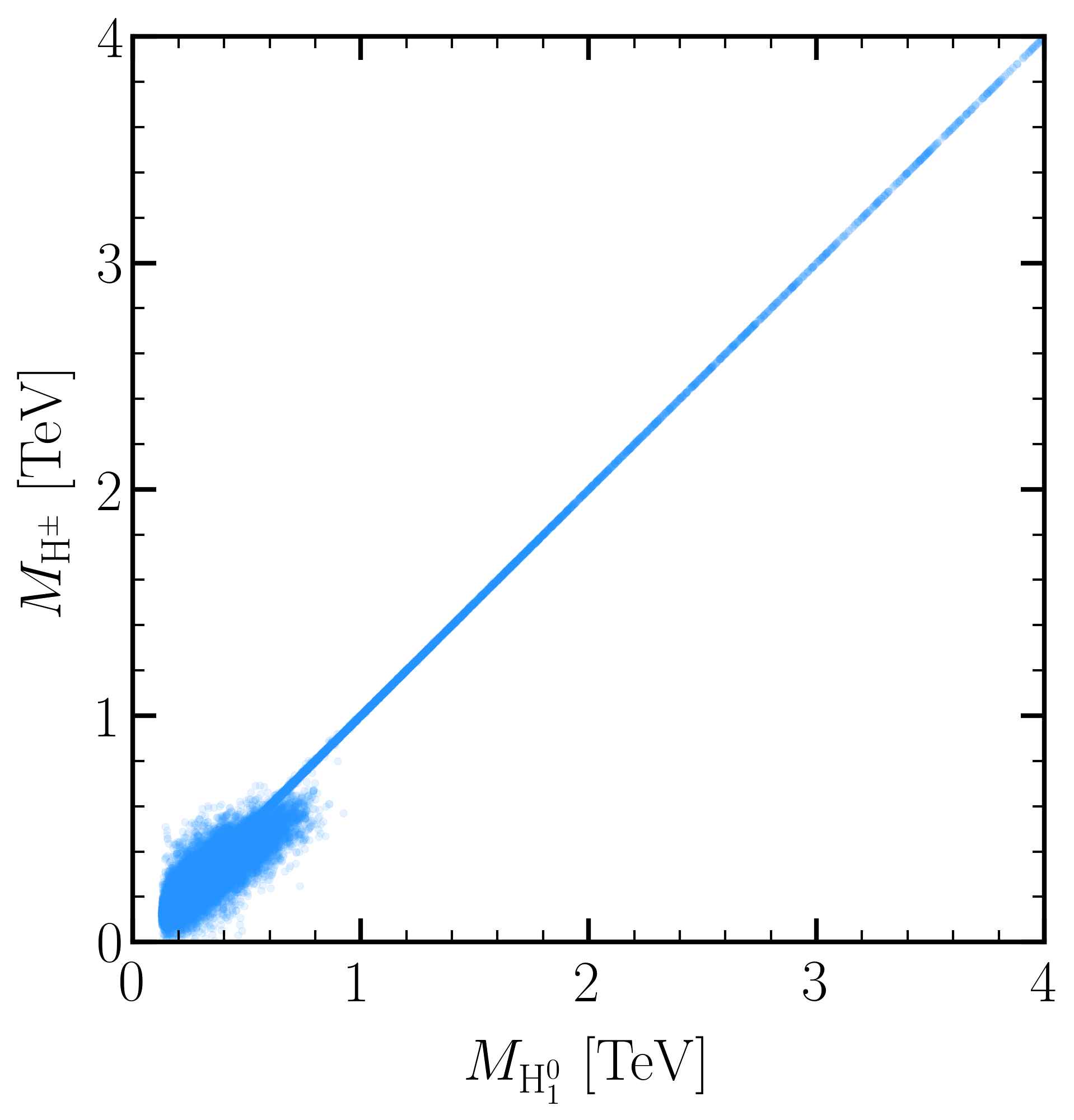}}
\caption{Masses of the new scalars, potentials with one minimum.\label{fig:vac1:Masses}}
\end{figure*}

\begin{figure*}[!ht]
\centering
\subfloat[$M_{\Fnf{2}}$ \vs\ $M_{\Fnf{1}}$.\label{sfig:vac2:Masses:M02vsM01}]{\includegraphics[width=0.44\textwidth]{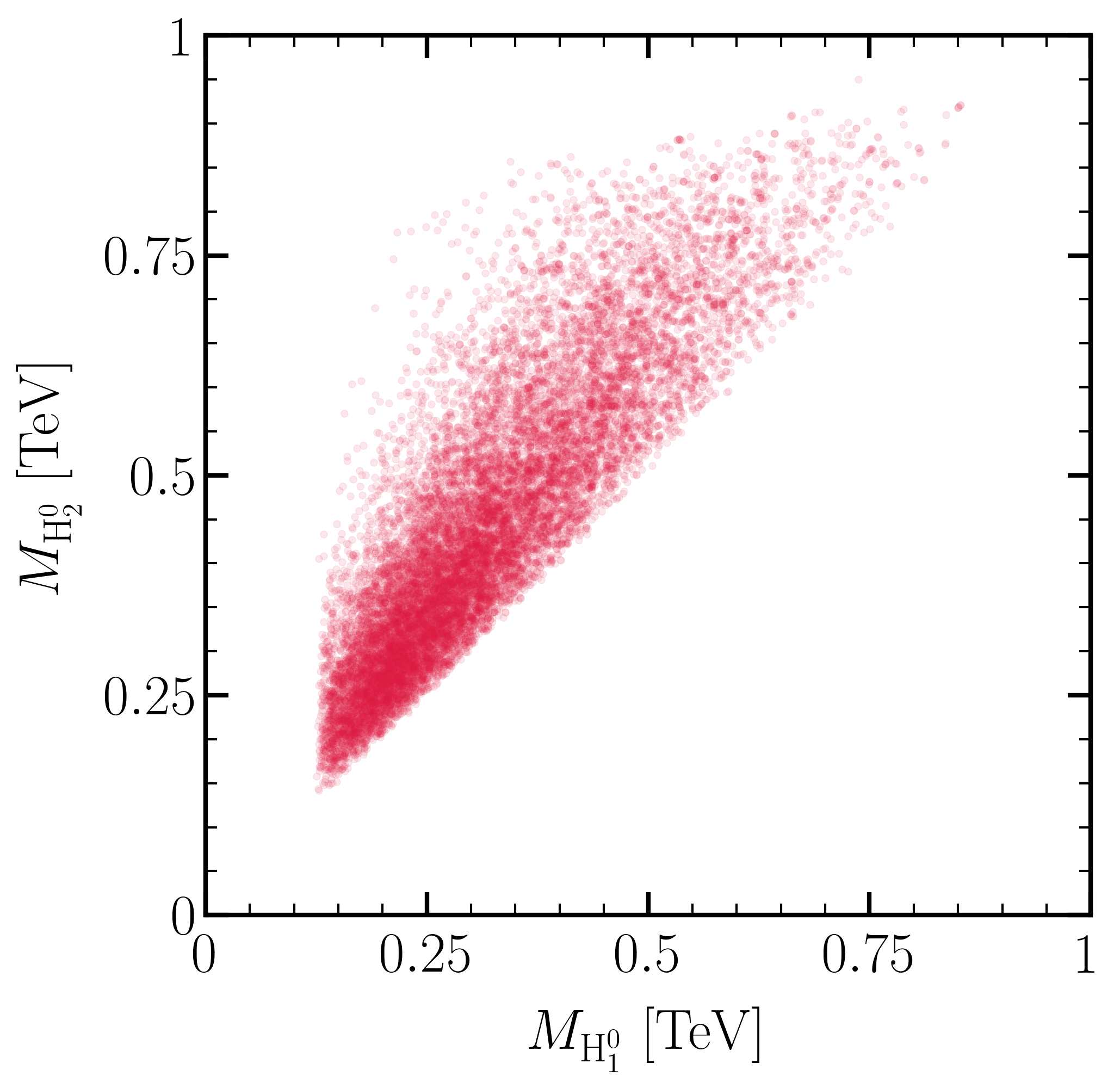}}\qquad
\subfloat[$M_{\FchfPM{}}$ \vs\ $M_{\Fnf{1}}$.\label{sfig:vac2:Masses:MC1vsM01}]{\includegraphics[width=0.44\textwidth]{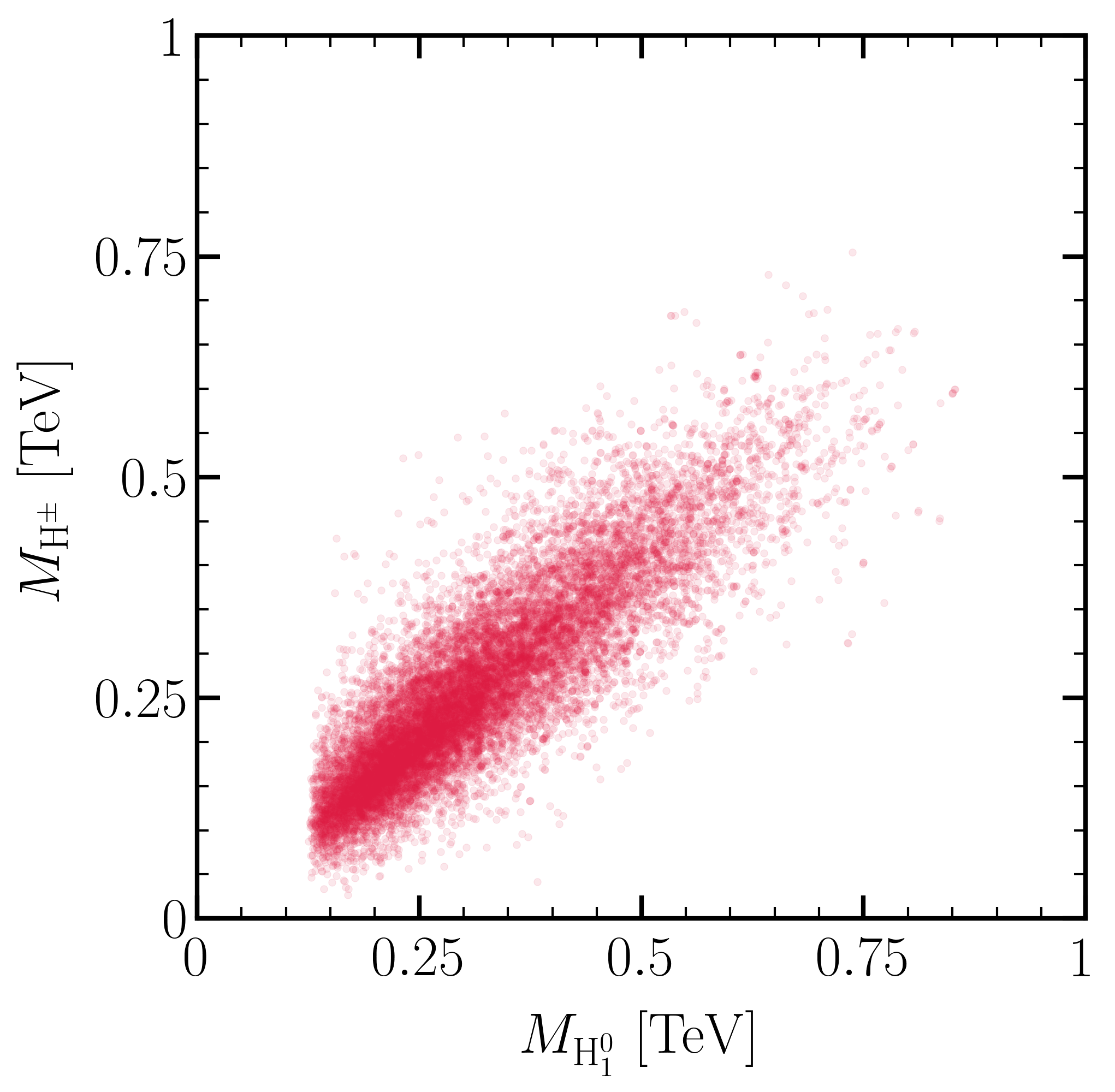}}
\caption{Masses of the new scalars, potentials with two minima.\label{fig:vac2:Masses}}
\end{figure*}

We now address a numerical analysis of the central question of this work: the goal is to explore the general 2HDM \emph{model space} focusing, for each model, on: (i) does the potential have one or two spontaneous symmetry breaking minima?, and (ii) in each case what is the spectrum of the new scalars? These are the steps followed in the process.
\begin{enumerate}
\item We generate random values of all quartic parameters and check whether the quartic part of the potential is bounded from below. This is achieved by minimizing and/or randomly probing that, for a fixed value $\vev{1}^2+\vev{2}^2\neq 0$, $\vevV_4(\vev{1},\vev{2},\vevPh{})\equiv\potV_4(\VEV{\SD{1}},\VEV{\SD{2}})>0$. If the set of generated $\QQ{}$'s does not fulfill the condition, it is rejected, and the process is restarted.
\item We generate random values of the quadratic parameters; $\vevV(\vev{1},\vev{2},\vevPh{})$ is now completely defined, and we proceed repeatedly (\eg\ 30 times) to minimize it numerically, with different random starting values of $\{\vev{1},\vev{2},\vevPh{}\}$.
\item The resulting list of minima is filtered, eliminating repeated minima (and potentials without spontaneous symmetry breaking). Specifically we compute the real vectors $\vec B=(\VEV{\SD{1}}^\dagger\VEV{\SD{1}},\VEV{\SD{2}}^\dagger\VEV{\SD{2}},\frac{1}{2}\VEV{\SD{1}}^\dagger\VEV{\SD{2}}+\frac{1}{2}\VEV{\SD{2}}^\dagger\VEV{\SD{1}},\frac{i}{2}\VEV{\SD{1}}^\dagger\VEV{\SD{2}}-\frac{i}{2}\VEV{\SD{2}}^\dagger\VEV{\SD{1}})$ for two candidate minima, $\vec B_{\rm Min\,1}$, $\vec B_{\rm Min\,2}$, and $\vec\Delta=\vec B_{\rm Min\,1}-\vec B_{\rm Min\,2}$: if $\vec\Delta\cdot\vec\Delta<$(Tolerance)$^2$, both candidate minima are ``the same'', and one is eliminated. We set, for example, Tolerance $= 10^{-5}$. Repeating this comparison, we are finally left with either one or two minima.
\item In case the potential has two minima, we choose the deepest one as the vacuum.
\item We now proceed with parameter redefinitions to obtain appropriate electroweak symmetry breaking vevs. For a vacuum with $\VEV{\SD{1}}=\frac{\vev{1,0}}{\sqrt 2}\left(\begin{smallmatrix}0\\ 1\end{smallmatrix}\right)$, $\VEV{\SD{2}}=\frac{\vev{2,0}e^{i\vevPh{0}}}{\sqrt 2}\left(\begin{smallmatrix}0\\ 1\end{smallmatrix}\right)$, all quadratic parameters are rescaled as
\begin{equation}
 \qq{a},\qq{ab}\ \mapsto\ \frac{\vev{\rm EW}^2}{\vev{0}^2}\,\qq{a},\qq{ab}\,,\quad \vev{0}^2\equiv\vev{1,0}^2+\vev{2,0}^2\,.
\end{equation}
The vacuum of the resulting potential is now $\VEV{\SD{1}}=\frac{\vev{1}}{\sqrt 2}\left(\begin{smallmatrix}0\\ 1\end{smallmatrix}\right)$, $\VEV{\SD{2}}=\frac{\vev{2}e^{i\vevPh{0}}}{\sqrt 2}\left(\begin{smallmatrix}0\\ 1\end{smallmatrix}\right)$, with $\vev{1}=\vev{\rm EW}\frac{\vev{1,0}}{\vev{0}}$ and $\vev{2}=\vev{\rm EW}\frac{\vev{2,0}}{\vev{0}}$: then $\vev{}=\vev{\rm EW}$ by construction.
\item We compute the mass matrices and their eigenvalues. As a check, there should be one null eigenvalue --- within numerical precision --- in each $\chMM$ and $\nMM$ corresponding to the would-be Goldstone bosons, and the remaining eigenvalues should be positive. With $m_0^2$ the smallest non-vanishing eigenvalue of $\nMM$, we now rescale all parameters
\begin{equation}
 \qq{a},\qq{ab},\QQ{x}\ \mapsto\ \frac{m_h^2}{m_0^2}\,\qq{a},\qq{ab},\QQ{x}\,,
\end{equation}
with $m_h=125.1$ GeV the Higgs mass. The vacuum is unchanged and now the lightest neutral scalar has mass $m_h$ (as already mentioned, this is in fact just a choice, but very appropriate in order to address the question of potentially heavy new scalars).
\item[] At this stage, we have a potential which is bounded from below, gives electroweak spontaneous symmetry breaking as desired, and has the lightest neutral scalar with the Higgs mass.
\item We now check that perturbative unitarity conditions, as discussed in \cite{Ginzburg:2005dt}, are fulfilled; if not, that potential is rejected and the process is restarted.
\item As additional checks, (i) two different parametrizations for the minimization of $\vevV(\vev{1},\vev{2},\vevPh{})$ are separately used, in terms of $\{\vev{1},\vev{2},\vevPh{}\}$ and in terms of $\{\vev{},\frac{\vev{2}}{\vev{1}},\theta\}$, and, also separately, (ii) a random unitary rotation of the $\{\SD{1},\SD{2}\}$ basis is performed before minimization (with all parameters in the potential accordingly transformed). The results of all three analyses are fully consistent.
\end{enumerate}
For the potentials thus obtained, we finally show the masses of the new scalars in \cref{fig:vac1:Masses} and \cref{fig:vac2:Masses}. Scatter plots are sufficient for our purpose --- we do not attach statistical meaning to the density/distribution of cases, our focus is on the ``allowed regions''. Figure~\ref{fig:vac1:Masses} corresponds to cases where the potential has a single minimum, while for \cref{fig:vac2:Masses} the potential has two minima. The plots speak for themselves: it is clear and remarkable that while in the case of a single minimum all the new scalars can have masses much larger than the electroweak scale (in a regime in which they are all degenerate), when the potential has two minima, all the new scalars have masses not larger than $\sim 1$ TeV, that is, the whole spectrum is bounded and there is no possible decoupling regime.


\begin{figure*}[!ht]
\centering
\subfloat[Potentials with one minimum.\label{sfig:vac1:v2v1vsMC1:gen}]{\includegraphics[width=0.44\textwidth]{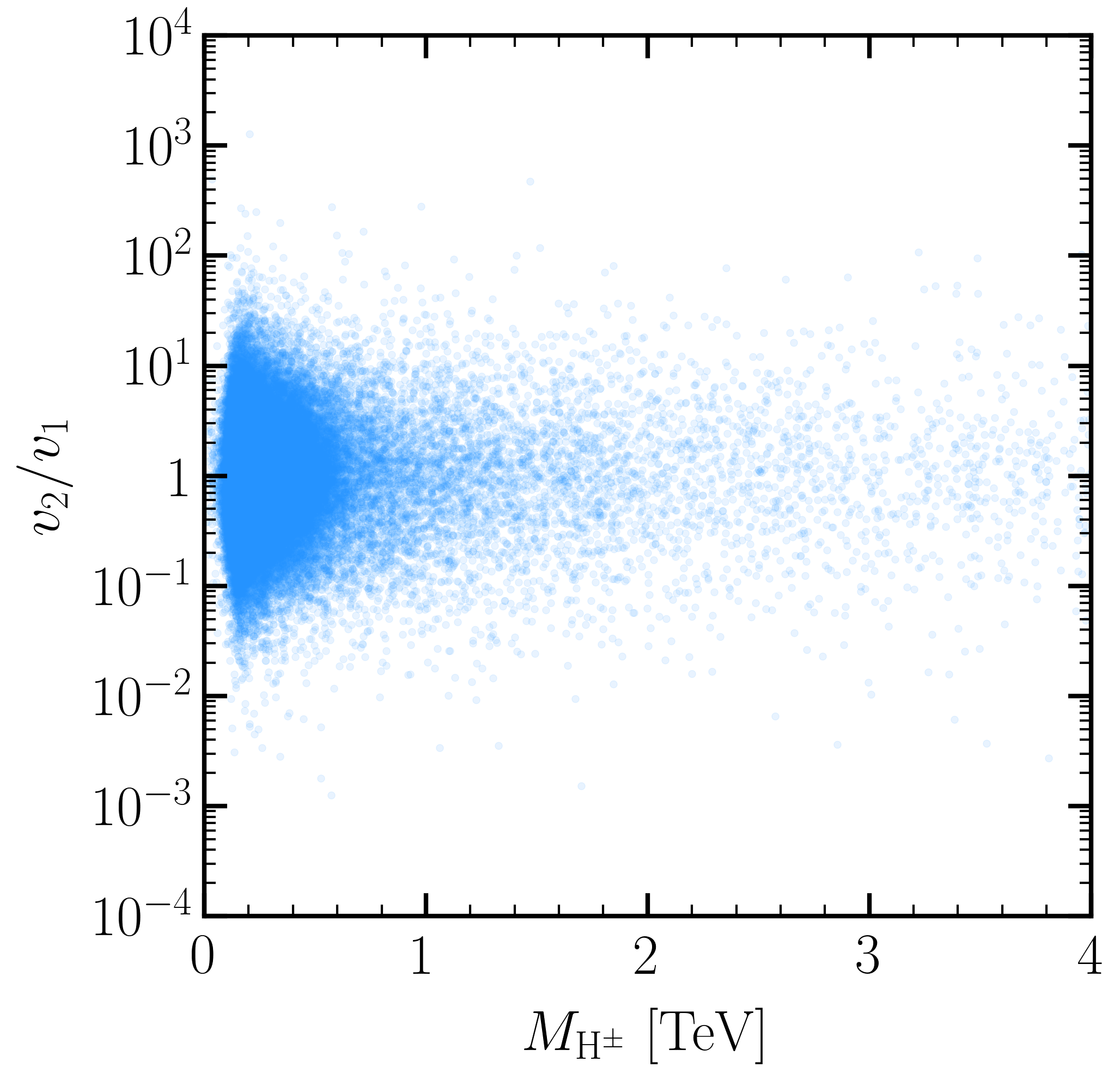}}\qquad
\subfloat[Potentials with two minima.\label{sfig:vac2:v2v1vsMC1:gen}]{\includegraphics[width=0.44\textwidth]{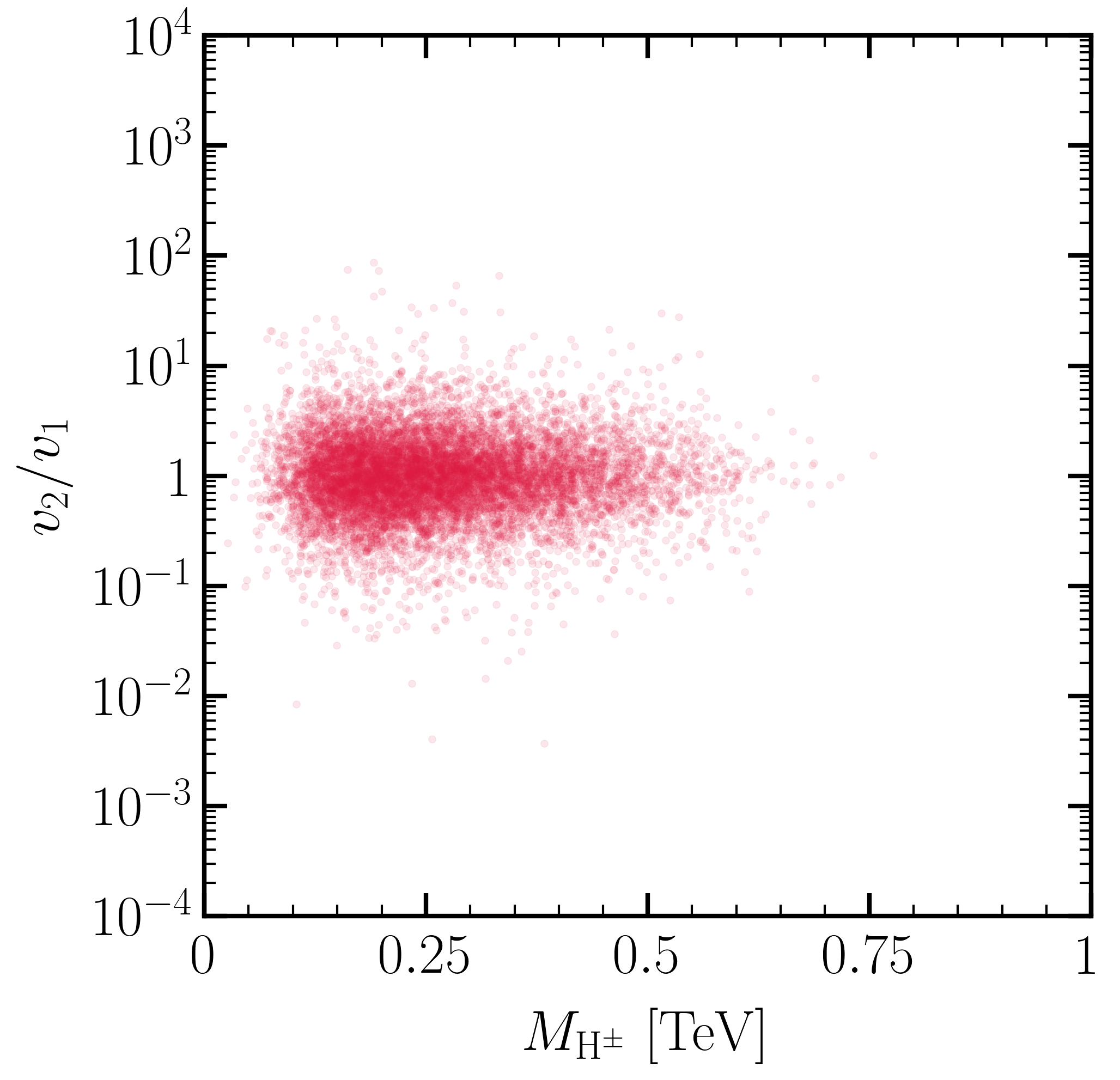}}
\caption{$\vev{2}/\vev{1}$ \vs\ $M_{\FchfPM{}}$, generic basis.\label{fig:v2v1vsMC1:gen}}
\end{figure*}

The first conclusion to draw is straightforward:
\begin{itemize}
 \item if the new scalars have masses larger than 1 TeV, then there is no additional minimum of the potential.
\end{itemize}
For new scalars with masses below 1 TeV, can one discriminate (at least to some extent) both possibilities? In order to provide an answer to that question, we first recall that in the parametrization of 2HDMs it is usual to employ the ratio of vevs $\tan\beta\equiv\frac{\vev{2}}{\vev{1}}$, with $\beta\in[0;\pi/2]$. Notice, however, that in general $\tan\beta$ has no physical meaning, since one can rotate arbitrarily in the space of $\{\SD{1},\SD{2}\}$ \cite{Haber:2006ue}. In \cref{fig:v2v1vsMC1:gen} we show $\frac{\vev{2}}{\vev{1}}$ \vs\ $M_{\FchfPM{}}$: no particular pattern appears to tell apart the two classes of potentials attending to $\frac{\vev{2}}{\vev{1}}$, as one could have expected.\footnote{The scarcity of points with $\frac{\vev{2}}{\vev{1}}<10^{-2}$ or with $\frac{\vev{2}}{\vev{1}}>10^{2}$ is just a consequence of the sampling and the logarithmic scale in \cref{fig:v2v1vsMC1:gen}: rotating the $\{\SD{1},\SD{2}\}$ basis with a uniform random angle, these regions amount to a fraction $\sim 10^{-2}$ of the available $\vev{2}/\vev{1}$ space.}

\begin{figure*}[!ht]
\centering
\subfloat[Potentials with one minimum.\label{sfig:vac1:v2v1vsMC1:diag}]{\includegraphics[width=0.44\textwidth]{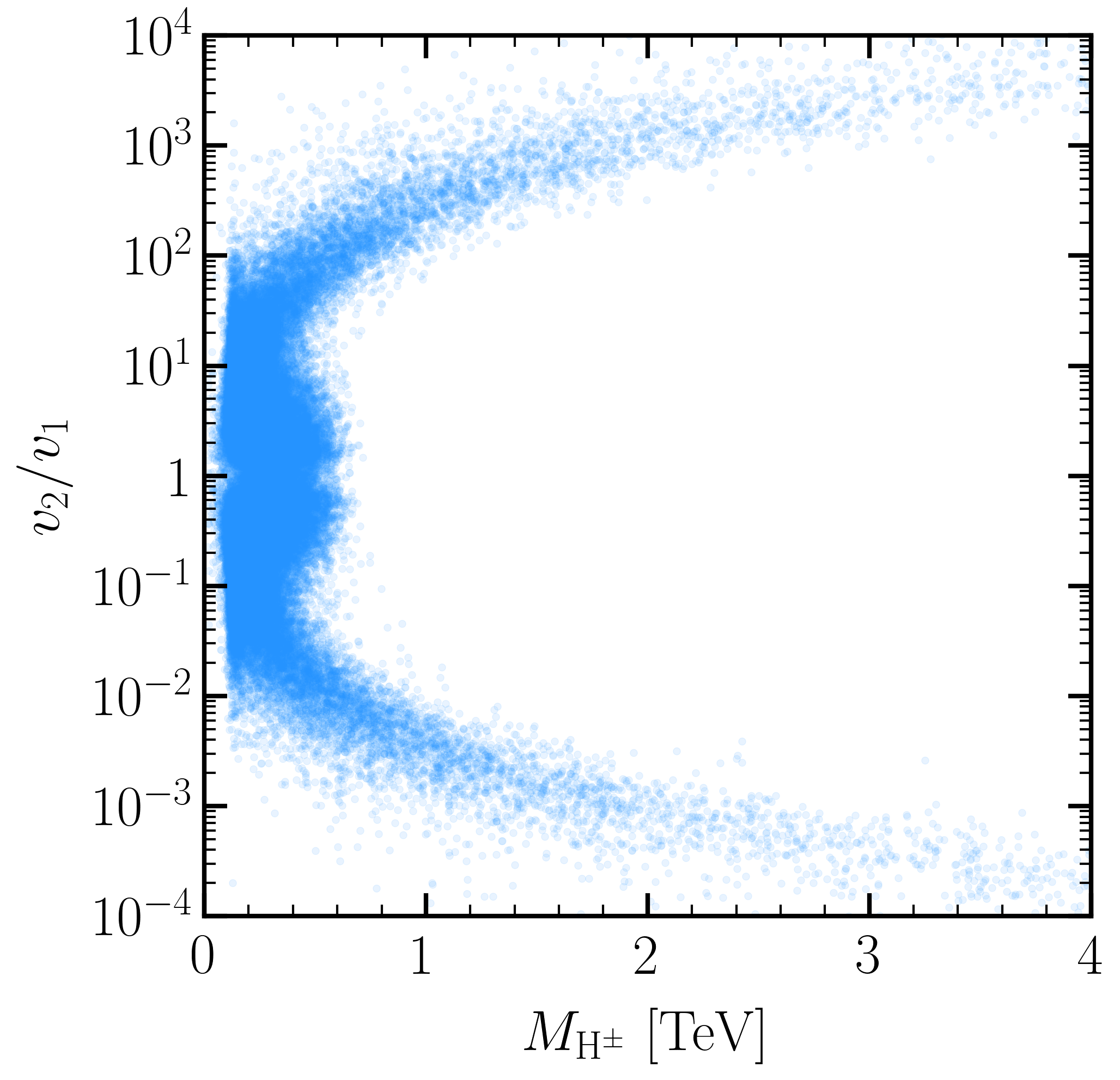}}\qquad
\subfloat[Potentials with two minima.\label{sfig:vac2:v2v1vsMC1:diag}]{\includegraphics[width=0.44\textwidth]{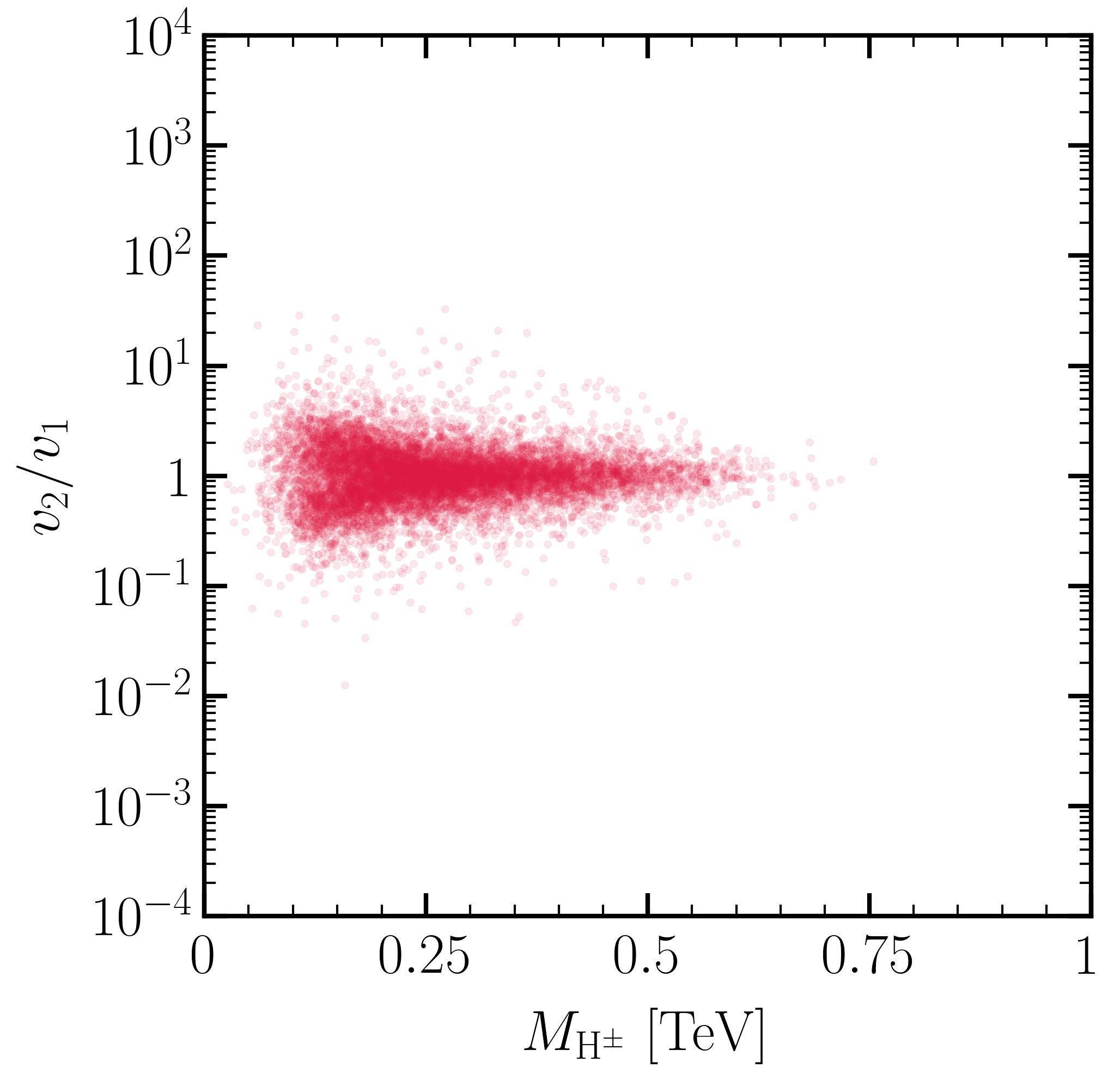}}
\caption{$\vev{2}/\vev{1}$ \vs\ $M_{\FchfPM{}}$, basis of diagonal quadratic couplings.\label{fig:v2v1vsMC1:diag}}
\end{figure*}

There is, however, a particular basis which is of interest. Considering the quadratic part of the potential in \cref{eq:2HDM:V:01},
\begin{equation}\label{eq:V2:matrix:01}
 \potV_2=\begin{pmatrix}\SDd{1} & \SDd{2}\end{pmatrix}\begin{pmatrix}\qq{1} & \qq{12}\\ \qqc{12} & \qq{2}\end{pmatrix}\begin{pmatrix}\SD{1}\\ \SD{2}\end{pmatrix}\,,
\end{equation}
one can introduce $(\SDd{1}, \SDd{2})=(\SDpd{1}, \SDpd{2})\,\mathcal U^\dagger$, with $\mathcal U\in U(2)$, such that
\begin{equation}
 \potV_2=\begin{pmatrix}\SDpd{1} & \SDpd{2}\end{pmatrix}\begin{pmatrix}\qq{\rm d1} & 0\\ 0 & \qq{\rm d2}\end{pmatrix}\begin{pmatrix}\SDp{1}\\ \SDp{2}\end{pmatrix}\,,
\end{equation}
that is, one can diagonalize the hermitian matrix of quadratic parameters in \cref{eq:V2:matrix:01} with $\qq{\rm d1,2}=\frac{1}{2}\left(\qq{1}+\qq{2}\pm\sqrt{(\qq{1}-\qq{2})^2+4|\qq{12}|^2}\right)$. We show in \cref{fig:v2v1vsMC1:diag} the ratio of vevs \vs\ $M_{\FchfPM{}}$ in that basis for both classes of potentials. Two relevant observations follow.
\begin{itemize}
\item The decoupling regime requires either very large or very small values of $\frac{\vev{2}}{\vev{1}}$.
\item Most interestingly, in the regime with masses below 1 TeV, values $\frac{\vev{2}}{\vev{1}}>10^2$ or $\frac{\vev{2}}{\vev{1}}<10^{-2}$ appear to distinguish between the cases of one and two minima. For values $10^{-2}<\frac{\vev{2}}{\vev{1}}<10^2$ one cannot establish that distinction.
\end{itemize}
We postpone further discussion to the next section.

\section{Discussion\label{SEC:Discussion}}
The central result to analyze, as illustrated with the numerical exploration of the previous section, is the fact that the masses of all the new scalars are necessarily bounded --- when perturbativity considerations constrain the quartic parameters --- if the scalar potential has 2 local minima. Starting with the chosen vacuum, given by $\{\vev{1},\vev{2}e^{i\vevPh{}}\}$, the stationarity conditions in \cref{eq:statcond:v1,eq:statcond:v2,eq:statcond:th} already fix 3 of the 4 quadratic parameters. The remaining freedom in the quadratic parameters, encoded in the combination $\qq{1}+\qq{2}$ or, equivalently, $\im{\qq{12}}$, could allow for a regime with masses of the new scalars much larger than $\vev{\rm EW}$. If there is a second minimum given by $\{\vev{1}^\prime,\vev{2}^\prime e^{i\vevPh{}^\prime}\}$, the stationarity conditions are also satisfied with $\vev{1}\mapsto\vev{1}^\prime$, $\vev{2}\mapsto\vev{2}^\prime$, $\vevPh{}\mapsto\vevPh{}^\prime$. Focusing on \cref{eq:statcond:th}, we thus have
\begin{multline}
-\im{\qq{12}e^{i\vevPh{}}}\vev{1}\vev{2}-\im{\QQ{5}e^{i2\vevPh{}}}\vev{1}^2\vev{2}^2\\
 -\im{(\QQ{6}\vev{1}^2+\QQ{7}\vev{2}^2)e^{i\vevPh{}}}\frac{\vev{1}\vev{2}}{2} =0\,
\end{multline}
and
\begin{multline}\label{eq:statcond:th:2vac}
-\im{\qq{12}e^{i\vevPh{}^\prime}}\vev{1}^\prime\vev{2}^\prime-\im{\QQ{5}e^{i2\vevPh{}^\prime}}\vev{1}^{\prime\,2}\vev{2}^{\prime\,2}\\
-\im{(\QQ{6}\vev{1}^{\prime\,2}+\QQ{7}\vev{2}^{\prime\,2})e^{i\vevPh{}^\prime}}\frac{\vev{1}^\prime\vev{2}^\prime}{2} =0\,.
\end{multline}
We assume $\vev{1}\vev{2}\neq 0$, $\vev{1}^\prime\vev{2}^\prime\neq 0$, and solve for $\re{\qq{12}}$ and $\im{\qq{12}}$:
\begin{align}
\re{\qq{12}}&=\frac{[E]\cos\vevPh{}^\prime-[E^\prime]\cos\vevPh{}}{\sin(\vevPh{}^\prime-\vevPh{})}\,,\\
\im{\qq{12}}&=\frac{[E^\prime]\sin\vevPh{}-[E]\sin\vevPh{}^\prime}{\sin(\vevPh{}^\prime-\vevPh{})}\,,
\end{align}
where
\begin{align}
[E]&=\im{\QQ{5}e^{i2\vevPh{}}}\vev{1}\vev{2}+\tfrac{1}{2}\im{(\QQ{6}\vev{1}^2+\QQ{7}\vev{2}^2)e^{i\vevPh{}}}\,,\\
[E^\prime]&=\im{\QQ{5}e^{i2\vevPh{}^\prime}}\vev{1}^\prime\vev{2}^\prime+\tfrac{1}{2}\im{(\QQ{6}\vev{1}^{\prime\,2}+\QQ{7}\vev{2}^{\prime\,2})e^{i\vevPh{}^\prime}}\,.
\end{align}
It is then clear that now \emph{all} quadratic parameters are bounded by perturbativity requirements on the $\lambda$'s, and thus the whole spectrum is bounded, as \cref{fig:vac2:Masses} illustrates transparently.

One caveat worth commenting on is the following: in \cref{eq:statcond:v1,eq:statcond:v2,eq:statcond:th}, the quadratic parameters appear multiplied by vevs and thus  strong hierarchies in these vevs, $\vev{1}\gg\vev{2}$ or $\vev{1}\ll\vev{2}$, might apparently invalidate the argument leading to boundedness of all quadratic parameters. This is, nevertheless, not the case: this potential concerns is bypassed by the fact that, working in a generic basis $\{\SD{1},\SD{2}\}$, one can always rotate to a different basis where such hierarchies are absent. Then, with quadratic parameters bounded in such a basis, no change of basis can make them arbitrarily large. The assumption $\vev{1}\vev{2}\neq 0$ and $\vev{1}^\prime\vev{2}^\prime\neq 0$ after \cref{eq:statcond:th:2vac} is justified on these same grounds.

The second point of interest concerns the result in \cref{fig:v2v1vsMC1:diag}, obtained in the basis where the hermitian matrix of quadratic parameters is diagonal, \cref{eq:V2:matrix:01}. Notice, incidentally, that spontaneous symmetry breaking requires that either $\qq{\rm d1}$, or $\qq{\rm d2}$, or both $\qq{\rm d1}$ and $\qq{\rm d2}$ are negative. If both are negative, with $\qq{1}+\qq{2}<0$ in \cref{eq:massCH:01,eq:massN:01}, then the whole spectrum is necessarily bounded (this might happen for a potential with either one or with two minima). Considering that specific basis leads to a conundrum: the number of quadratic parameters is reduced to 2, while there are 3 stationarity conditions. How can we have mass spectra that are not bounded, since it might appear that all quadratic parameters can be expressed as quartic parameters times vevs? --- the spectrum of masses does not depend, of course, on the basis. The point is subtle: in that special basis, the stationarity conditions read
\begin{align}
\label{eq:statcond:spec:v1}
\partial_{\vev{1}}\vevV =& \qq{\rm d1}\vev{1}+\QQ{1}\vev{1}^3+\frac{1}{2}(\QQ{3}+\QQ{4})\vev{1}\vev{2}^2
+\re{\QQ{5}e^{i2\vevPh{}}}\vev{1}\vev{2}^2 \nonumber\\
&\ +\frac{3}{2}\re{\QQ{6}e^{i\vevPh{}}}\vev{1}^2\vev{2}+\frac{1}{2}\re{\QQ{7}e^{i\vevPh{}}}\vev{2}^3= 0 \,,\\
\label{eq:statcond:spec:v2}
\partial_{\vev{2}}\vevV =& \qq{\rm d2}\vev{2}+\QQ{2}\vev{2}^3+\frac{1}{2}(\QQ{3}+\QQ{4})\vev{1}^2\vev{2}
+\re{\QQ{5}e^{i2\vevPh{}}}\vev{1}^2\vev{2} \nonumber\\
&\ +\frac{1}{2}\re{\QQ{6}e^{i\vevPh{}}}\vev{1}^3+\frac{3}{2}\re{\QQ{7}e^{i\vevPh{}}}\vev{1}\vev{2}^2=0 \,,\\
\label{eq:statcond:spec:th}
\partial_{\vevPh{}}\vevV =& -\im{\QQ{5}e^{i2\vevPh{}}}\vev{1}^2\vev{2}^2-\im{(\QQ{6}\vev{1}^2+\QQ{7}\vev{2}^2)e^{i\vevPh{}}}\frac{\vev{1}\vev{2}}{2} \nonumber\\
&=0 \,.
\end{align}
The first aspect to notice is that the last equation, $\partial_{\vevPh{}}\vevV=0$, does not involve $\qq{\rm d1}$ and $\qq{\rm d2}$. There are just 2 equations for the 2 quadratic parameters and one can straightforwardly obtain
\begin{equation}
\begin{aligned}
\qq{\rm d1}+\qq{\rm d2} = &-\QQ{1}\vev{1}^2-\QQ{2}\vev{2}^2-\frac{1}{2}\left(\QQ{3}+\QQ{4}+2\re{\QQ{5}e^{i2\vevPh{}}}\right)\vev{}^2\\
&-\re{\QQ{6}e^{i\vevPh{}}}\frac{\vev{1}}{2\vev{2}}(3\vev{2}^2+\vev{1}^2)\\
&-\re{\QQ{7}e^{i\vevPh{}}}\frac{\vev{2}}{2\vev{1}}(3\vev{1}^2+\vev{2}^2)\,.
\end{aligned} 
\end{equation}
It is then clear that $\qq{\rm d1}+\qq{\rm d2}\gg\vev{\rm EW}^2$ is still feasible provided there is a strong hierarchy in the vevs. This is exactly the behaviour that \cref{fig:v2v1vsMC1:diag} shows for $M_{\FchfPM{}}> 1$ TeV. One can then interpret the distinction between one and two minima through $\vev{2}/\vev{1}$ for $M_{\FchfPM{}}< 1$ TeV as a continuation of the just-one-minimum case with milder vev hierarchies and $\qq{\rm d1}+\qq{\rm d2}\sim \vev{\rm EW}^2$.

To close this discussion, some comments concerning "the Higgs basis" are of interest. The "Higgs basis", mentioned in connection with the diagonalization of the charged mass matrix in \cref{eq:CHMass:HB}, is a basis $\{H_{1},H_{2}\}$ such that $\langle H_2\rangle =(0,0)^T$ (it is not unique, it is defined up to rephasings of $H_{2}$). The Higgs basis is particularly appropriate to relate parameters of the potential with physical quantities. With $\SD{1}\to H_1$, $\SD{2}\to H_2$, and $\qq{1}\to Y_1$, $\qq{2}\to Y_2$, $\qq{12}\to Y_3$ and $\QQ{j}\to Z_j$ the parameters in the Higgs basis, the stationarity conditions \cref{eq:statcond:v1,eq:statcond:v2,eq:statcond:th} are reduced in the Higgs basis to
\begin{equation}\label{eq:stat:HiggsBasis:01}
0=Y_1\vev{}+Z_1\vev{}^3,\qquad 0=\re{Y_3}\vev{}+\frac{1}{2}\re{Z_6}\vev{}^3\,.
\end{equation}
One can simply read \cref{eq:statcond:v1,eq:statcond:v2,eq:statcond:th} with parameters in the Higgs basis and set $\vev{1}\to\vev{}$ and $\vev{2},\vevPh{}\to 0$, to obtain \cref{eq:stat:HiggsBasis:01}. Notice that $Y_2$ is absent from \cref{eq:stat:HiggsBasis:01} and, furthermore,
\begin{align}
\TR{\chMM}&=M_{\FchfPM{}}^2\nonumber\\ &=Y_1+Y_2+\vev{}^2(Z_1+Z_3/2)=Y_2+\frac{\vev{}^2}{2}Z_3\,,\\
\TR{\nMM}&=m_h^2+M_{\Fnf{1}}^2+M_{\Fnf{2}}^2\nonumber\\ &=2(Y_1+Y_2)+\vev{}^2(4Z_1+Z_3+Z_4)\nonumber\\ &=2Y_2+\vev{}^2(2Z_1+Z_3+Z_4)\,.
\end{align}
A decoupling regime for the new scalars requires $Y_2\gg \vev{}^2$. What prevents that possibility when the potential has two minima? The answer is straightforward, besides the usual Higgs basis solution of the stationarity conditions in \cref{eq:stat:HiggsBasis:01}, there is another solution with $\langle H_2\rangle \neq (0,0)^T$: it simply corresponds to \cref{eq:statcond:v1,eq:statcond:v2,eq:statcond:th} with the parameters in the Higgs basis. The point is that $Y_2$ is then constrained by \cref{eq:statcond:v2}, and one cannot have $Y_2\gg \vev{}^2$.
Finally, concerning \cref{fig:v2v1vsMC1:diag} and the discussion on the ratio of vevs $\tan\beta=\vev{2}/\vev{1}$ \emph{in the basis of diagonal quadratic couplings}, notice that, in terms of the Higgs basis parameters, one has
\begin{equation}
\tan 2\beta=2\frac{\abs{Y_3}}{Y_2-Y_1}\,.
\end{equation}
One can readily observe that a decoupling regime of the new scalars with $Y_2\gg \vev{}^2$ and $\abs{Y_1},\abs{Y_3}$ bounded by \cref{eq:stat:HiggsBasis:01} implies $\tan 2\beta\simeq 0$, that is $\beta\simeq 0,\pi/2$, in full agreement with \cref{sfig:vac1:v2v1vsMC1:diag}.

Some final comments are in order. Typical usage of 2HDMs involves setting $\vev{}=\vev{\rm EW}$, choosing parameters such as $\tan\beta=\vev{2}/\vev{1}$, masses and mixings, and then enforcing through the stationarity conditions and the mass matrices what the corresponding potential should be. The results of this work might be helpful to dismiss or not the concerns that the existence of a second minimum of the potential can trigger, such as, for example, if the considered vacuum is the lowest one, or the cosmological consequences of having domains with different electroweak symmetry breaking properties. If one is considering a 2HDM with new scalars heavier than 1 TeV, there is no concern in that respect. For new scalars lighter than 1 TeV, if $\vev{2}/\vev{1}$ (in the basis of diagonal quadratic terms) is not within the range $[10^{-2};10^2]$, there is again no concern in that respect, the potential has no additional vacuum. By contrast, for $\vev{2}/\vev{1}\in[10^{-2};10^2]$ the scalar potential can have a second minimum, and it might not be safely ignored. It is relevant to stress that although \cref{fig:v2v1vsMC1:diag} is obtained under the assumption that the vacuum is the deeper minimum when two minima are present, this aspect holds on the same grounds in case the other minimum is chosen as the vacuum: expanding fields around the other minimum, the masses are also bounded, and the ratio of vevs (in the basis of diagonal quadratic terms) of that shallower minimum is in the range $[10^{-2};10^{2}]$. This regime with light scalars might give signals within the discovery reach of the LHC, and could be of immediate interest. To address such phenomenological implications, the Yukawa couplings to fermions cannot be ignored: they are thus beyond the scope of this work, which focuses on the scalar sector on its own.

\section{Conclusions\label{SEC:Conclusions}}
We have shown how a simple, global property of the scalar potential in general 2HDMs, namely the existence of two local minima, can have a clear and somehow unexpected consequence: the whole scalar spectrum is bounded, with all masses below 1 TeV, owing to perturbativity constraints. In addition, we have explored the possibility of distinguishing, to some extent, the presence of two minima through the ratio of vevs in a specific basis.

\section*{Acknowledgements}
The authors thank J. Silva for comments on the manuscript. The authors acknowledge support from Spanish MICIU/AEI/10.13039/501100011033/ through grant PID2023-151418NB-I00 and the \emph{Severo Ochoa} project CEX2023-001292-S. \textit{Conselleria de Innovaci\'on, Universidades, Ciencia y Sociedad Digital} from \textit{Generalitat Valenciana} (Spain) and \emph{Fondo Social Europeo} support JMC, MN and TT through projects CIDEGENT/2019/024 and CIESGT/2024/21.

\bibliographystyle{elsarticle-num}
\bibliography{final-2HDM-vac}

\end{document}